\documentclass[aps,pre,superscriptaddress,showpacs]{revtex4-1}
\usepackage{graphicx,amsbsy,amsmath,epsfig,natbib,float}
\usepackage{amssymb,latexsym,wasysym,pifont,mathrsfs}
\usepackage{comment,verbatim,multirow,array,sidecap,color}
\usepackage{lineno,setspace,soul,cancel}
\usepackage[normalem]{ulem}

\begin{document}

\title{Mechanical Behaviour of Glasses and Amorphous Materials}

\author{Anshul D. S. Parmar}\affiliation{School of Advanced Materials, Jawaharlal
  Nehru Center for Advanced Scientific Research, Jakkur Campus, Bengaluru 560064, India.}
\author{Srikanth Sastry}  \email{sastry@jncasr.ac.in\\
  Advances in the Chemistry and Physics of Materials: Overview of Selected Topics, World Scientific (2020)}

\begin{abstract}
A wide range of materials can exist in microscopically disordered solid forms, referred to as amorphous solids or glasses. Such materials -- oxide glasses and metallic glasses, to polymer glasses, and soft solids such as colloidal glasses, emulsions and granular packings --  are useful as structural materials in a variety of contexts. Their deformation and flow behaviour is relevant for many others.  Apart from fundamental questions associated with the formation of these solids, comprehending their mechanical behaviour is thus of interest, and of significance for their use as materials. In particular, the nature of plasticity and yielding behaviour in amorphous solids has been actively investigated. Different amorphous solids exhibit behaviour that is apparently diverse and qualitatively different from those of crystalline materials. A goal of recent investigations has been to comprehend the unifying characteristics of amorphous plasticity and to understand the apparent differences among them. We summarise some of the recent progress in this direction. We  focus on insights obtained from computer simulation studies, and in particular those employing oscillatory shear deformation of model glasses. 
\end{abstract}
\maketitle 

\section{Introduction}
%2-3 pages 

Amorphous materials are ubiquitous in nature, and arise as among the common solid forms of a wide range of substances. The most familiar are various forms of silica glass, but chalcogenide glasses and amorphous metallic alloys or metallic glasses\cite{ramamurty} are other common examples of what are some times termed {\it hard glasses}. At the other end of the spectrum are a variety of {\it soft materials}, which superficially form a very distinct class, but share some  fundamental characteristics with  hard glasses\cite{Bonn2017c,Nicolas2018}. Some examples are dense colloidal assemblies, gels, emulsions, pastes and foams. These materials are typically studied for their flow or rheological properties, but many of them are observed to possess a {\it yield stress} below which they exhibit solid-like response and above which they exhibit viscous flow. Other common materials that may deserve mention are polymer glasses and granular materials. Many biological and geophysical formations on relevant scales are well described as amorphous solids, and their analysis is therefore informed by the understanding of amorphous solids (statistics of earthquakes is a familiar and oft-quoted example).

\begin{figure}
\centering
  \includegraphics[scale=0.12]{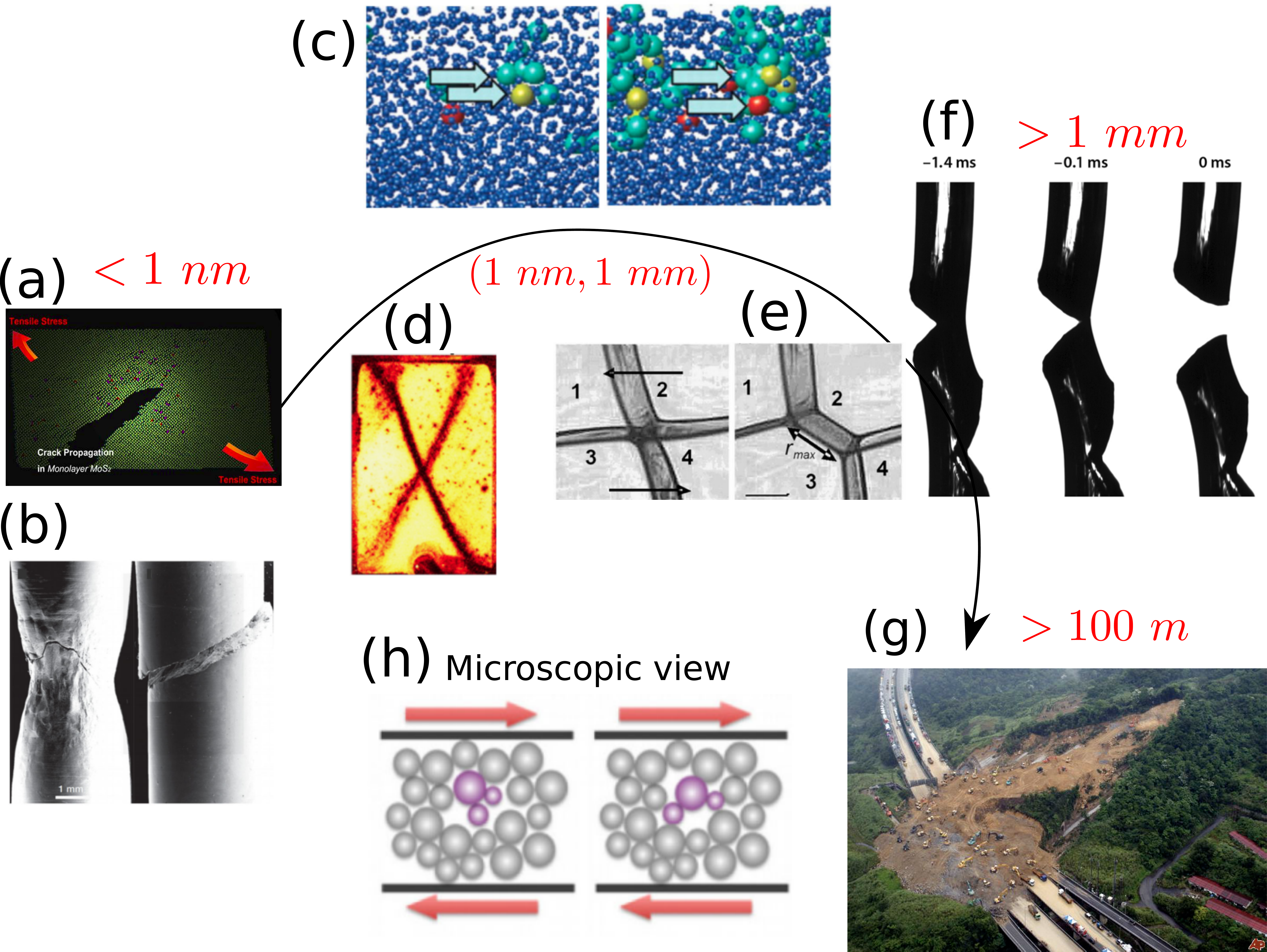}
  \caption{Examples of amorphous solids and deformations. (a) Image of a crack with atomic resolution in  a planar $MoS_2$ sample \cite{ly2017dynamical}. (b) Bulk metallic glass pillars under uniaxial stress.  The yield behaviour of a composite glass reinforced with dendrites becomes more ductile \cite{hofmann2008designing}. 
  (c) The plastic events in a slowly sheared colloidal system is represented in terms of a change in the neighbours of colloids. The color map and the size scale  represent the number of nearest neighbors lost during a plastic event. \cite{schall2007structural}.
  (d) The strain localisation in a granular pack of micrometer size glass beads under biaxial compression \cite{le2014emergence}.
  (e) Localized rearrangements in a bubble raft. The interface between neighboring bubbles exhibits a $T1$ event, triggered by applied strain \cite{biance2009topological}. 
  (f) The stretching of a polymer melt. At sufficiently high strain rates the polymer melt behaves in a way similar to a solid \cite{huang2016multiple}.
  (g) Landslide hit highway in Malaysia (source: China Press, The Straits Times). 
  (h) The mechanical response to applied deformation is illustrated to involve local rearrangements.  \cite{bocquet2009kinetic}. }
  \label{fig:def_am_solids}
\end{figure}

A common feature of these substances is the lack of microscopic order, unlike crystalline materials. In terms of their mechanical and flow behaviour, all these materials exhibit solid-like behaviour, in that upon the application of small external stresses, they respond elastically, with deformations being proportional to the applied stresses, and reversible, so that they return to the original state when the external stresses are removed. At larger applied stresses, they display {\it plastic} deformation, which remain when the applied stresses are removed. The point of onset of plasticity, the yield point, is thus an important material characteristic, which precedes various forms of failure of these materials. This broad and simplified description of mechanical behaviour is superficially the same as that for crystalline materials, and thus one must clarify what distinguishes amorphous materials from crystalline solids. A key difference is the translational order present in crystals, and the consequent possibility to have well defined structural defects. In particular, the plastic response of crystals is described in terms of the presence and movement of dislocation defects.  In amorphous solids, there is no translational order, and thus no well defined notion of defects that are instrumental in plastic deformation. A related key distinction is that since amorphous solids are typically obtained as preparation protocol dependent, out of equilibrium, materials, the amorphous state is not unique for a given substance. The preparation method plays a key role in their observed behaviour, and a modification of the structure is often a key element in the response to applied stresses (see Figure \ref{fig:def_am_solids}) \cite{argon1979plastic,Schuh2007a,Schall2010,Falk2011,Barrat2011,Bonn2017c,Nicolas2018}.

Studies of mechanical behaviour of amorphous solids focus on these key distinctions from the crystalline state, but it may be noted that the description of crystalline solids on large scales shares some of these features \cite{sethna2017deformation}. A second important general consideration pertains to the notion of solidity of either crystalline or amorphous materials. On general grounds, it has been argued that a solid at finite temperature, under finite external stresses, will always deform and flow at finite shear rates and the distinction between solids with a finite shear modulus and viscoelastic fluids must be made with some care \cite{sausset2010solids,yoshinomezard}.
%{\color{red} In the recent theoretical framework, the rigidity is considered to be valid only is the metastable state of finite systm. The existence of the rigidity is further correlated with slow collective processes e.g., particle's rearrangements, breaking this symmetry leads to a first-order transition and phase coexistence between a rigid solid and plasticity \cite{nath2018existence}.}
Thus, many attempts to understand yielding as  a well defined transition focus on the limit of zero temperature, as discussed below. On the other hand, an understanding of behaviour in limiting cases forms the basis for describing finite temperature and shear rate behaviour \cite{nath2018existence}.

%for real materials, it is both necessary and possible to discuss such a change in behaviour within reasonable windows of time scales. 

\begin{figure}
\centering
  \includegraphics[scale=0.35]{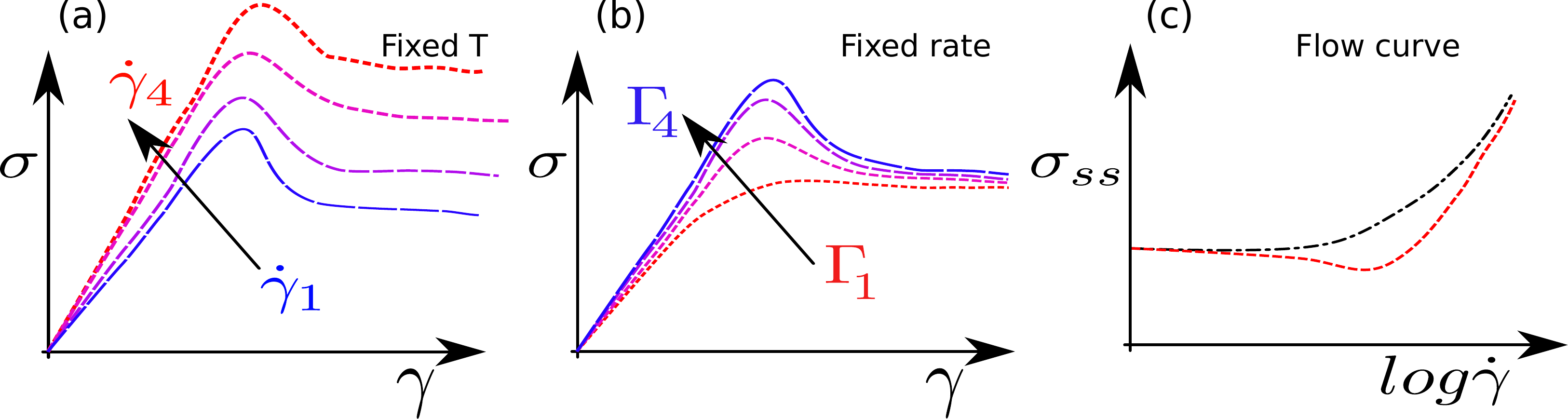}
  \caption{An illustrative image showing stress strain curves with and without overshoots, and flow curves. (a) A sketch of the stress-strain curves for a glass  sheared at different strain rates. Larger rates ($\dot\gamma_{1}<\dot\gamma_{2}<\dot\gamma_{3}<\dot\gamma_{4}<...$) lead to larger yield stresses and larger steady state stresses \cite{Falk2011}.
   (b) A sketch of stress-strain curves for glasses prepared at different cooling rates.    The lower cooling rates ($\Gamma_{1}>\Gamma_{2}>\Gamma_{3}>\Gamma_{4}...$) result in better annealed glasses. More annealed glasses show  larger maximum stresses.  (c) Steady-state flow curves, i.e., dependence of the steady-state shear stress $\sigma_{ss}$ on the shear rate $\dot\gamma$ for a yield stress fluid. Homogeneous flow corresponds to a monotonic flow curve, of the Herschel-Bulkley form.  \cite{herschel1926konsistenzmessungen}. A regime with a negative slope in the flow curve is generally associated with instability and the presence of shear bands \cite{coussot2010physical}.}
  \label{fig:strain_strain_strainrates}
\end{figure}

\begin{figure}
  \includegraphics[scale=0.15]{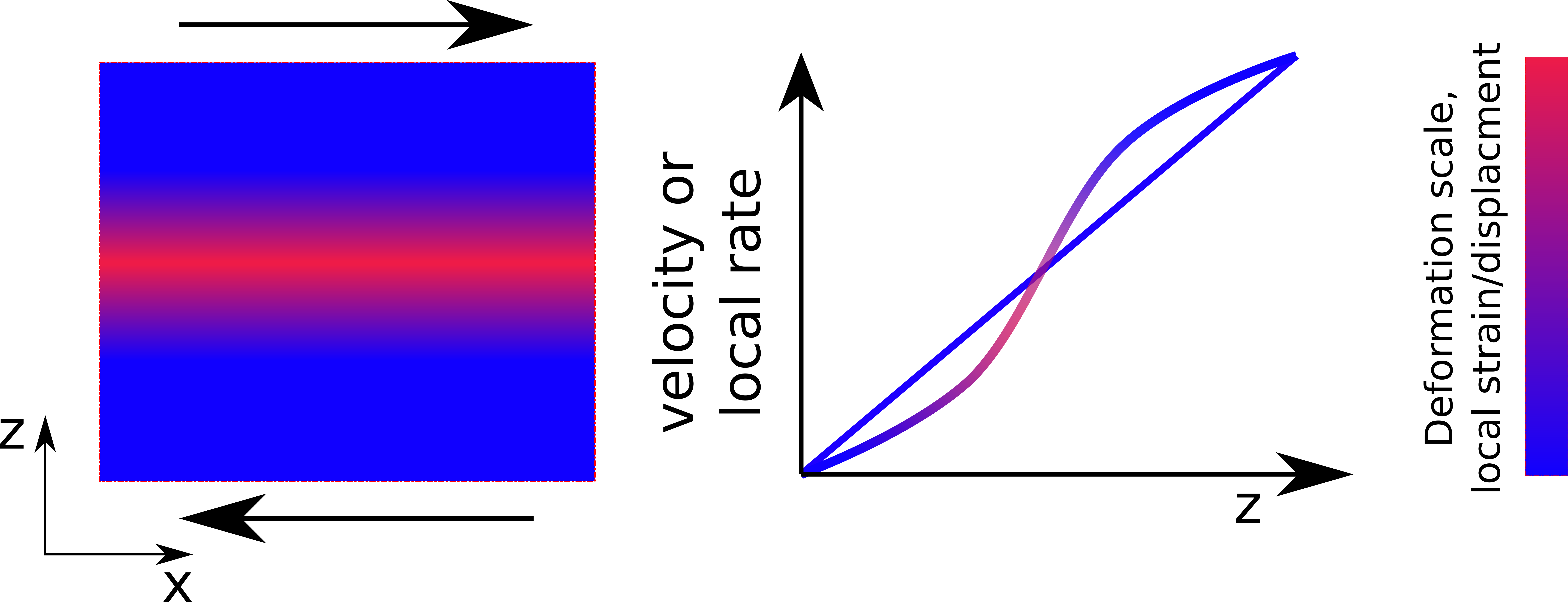}
  \caption{Illustrative images illustrating (left) strain localization and (right) shear banding under flow. (left) The blue region corresponds to lower strain and the red regime corresponds to localization of strain. (right) The velocity profile in a shear banding system, compared with uniform flow.}
  \label{fig:strainlocalisation}
\end{figure}

Figure \ref{fig:strain_strain_strainrates} shows an illustration of stress-strain curves of amorphous solids, displaying some typical behaviours observed. In some materials, the dependence of stress on the strain is gradual, and smoothly approaches an asymptotic steady flow value. In other cases, one observes a stress overshoot, followed by a drop, before approaching the steady flow value. It is typically observed that in the cases where one observes a stress overshoot, yielding is accompanied by the presence of strain localisation, {\it i. e.}, the spatial localisation of plastic strain within narrow regions of the sample (illustrated in Figure \ref{fig:strainlocalisation} (a)). Such localisation of plastic strain is, for example, observed in metallic glasses, and leads to brittle failure (Figure \ref{fig:strainlocalisation} (b)). The stress overshoot and the flow stress also depend on the rate at which strain is applied. The yielding and flow behaviour can also be represented in terms of steady state flow stresses as a function of strain rates, as shown in Figure \ref{fig:strain_strain_strainrates} (c).  The steady flow stresses typically increase with strain rates 
but in some cases, the flow curves are non-monotonic  (illustrated in Figure \ref{fig:strain_strain_strainrates} (c)). The non-monotonic flow curves are observed to correspond to the presence of {\it shear bands} within which the bulk of the flow becomes localised (illustrated in Figure \ref{fig:strainlocalisation})\cite{coussot2010physical}. The presence of strain localisation and shear banding represent a key feature of yielding behaviour that requires explanation.

Thus, some of the key questions to address in discussing the mechanical behaviour of amorphous solids pertains to the nature of elementary processes of plasticity that occur in amorphous solids, the nature of interactions among them, the nature of the yielding transition, the dependence of the mechanical response on the preparation history and structural properties of the amorphous solids, the nature of structural change during deformation, and the presence or otherwise of strain localisation and shear banding. We address these questions in this review, providing a summary of progress in answering them. Only some selected themes are summarised, and many omitted, to permit an adequate discussion of the topics chosen. Hence, the overview presented should be treated more as a perspective than a comprehensive review. In Section 2, we provide a general background and overview of theoretical and computational approaches to develop a description of mechanical behaviour and yielding in amorphous solids. We do not discuss relevant experimental results specifically, but these are well summarised in various reviews we mention \cite{ramamurty,Bonn2017c,Nicolas2018}. Section 3 discusses selected recent results, that address the nature of the yielding transition, the influence of sample preparation and annealing, and shear banding. Section 4 contains a summary of current status and outlook for further investigations. 

\section{Background and Previous Work} 
%5 - 6 pages 

%SS -- What to include? 
%SS -- Eshelby, STZ, STZ theory, interacting Eshelby's, mElastoplastic models, RFIM, Avalanches  

% Nature of elementary events of plastic reorganization 

\begin{figure}[ht!]
\centering
  \includegraphics[scale=0.25]{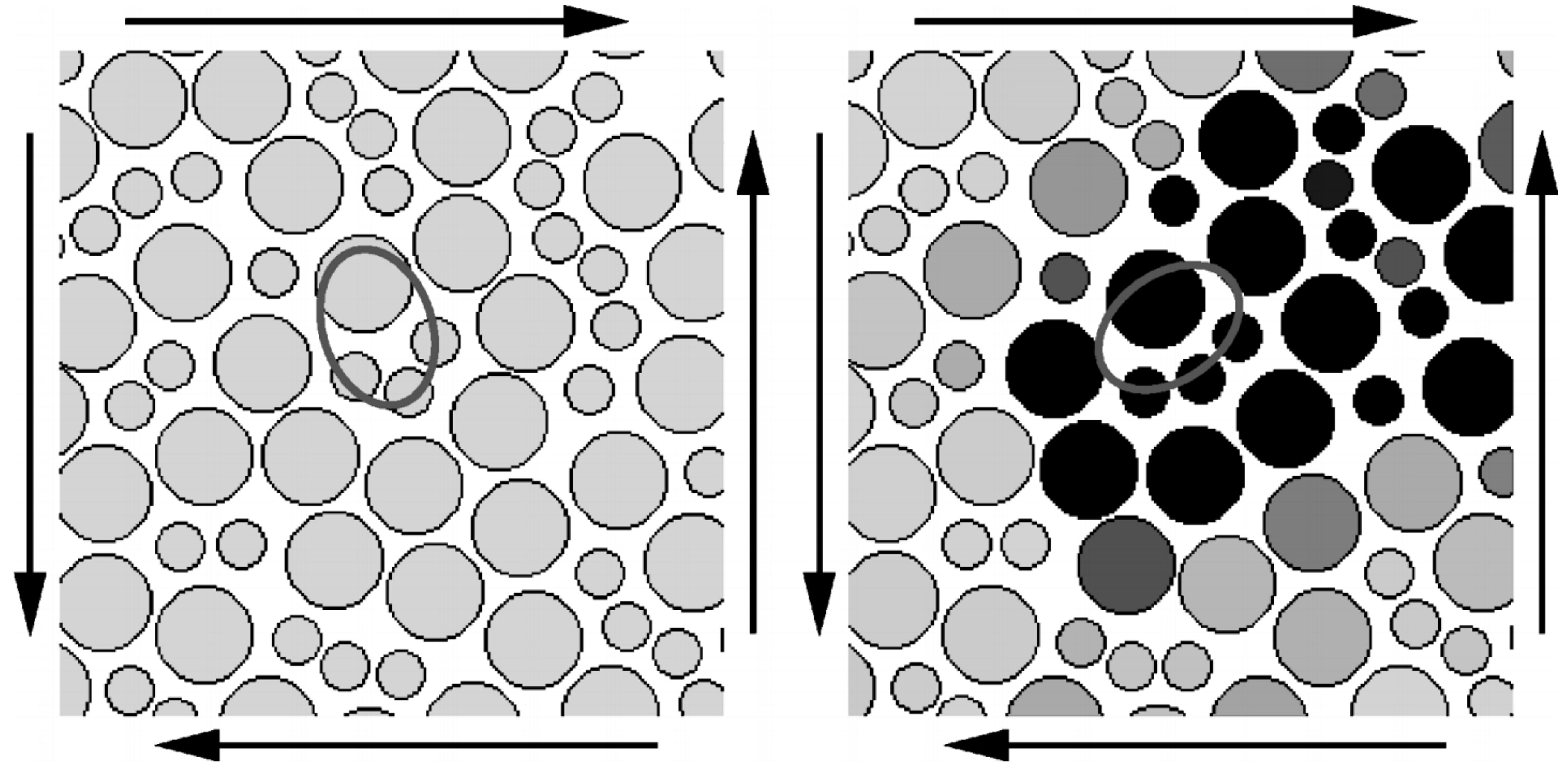}
  \caption{An illustration of plastic rearrangements at the particle scale: Particle positions  before and after a shear transformation event. The darker area represents the region that deviated from the affine deformation field. During this event, the indicated cluster of one large and three small particles changes orientation. (source \cite{falk1998dynamics})}
  \label{fig:STZ}
\end{figure}

The nature of rearrangements leading to plastic deformation have been investigated over several years, as a step in developing a description of plasticity in amorphous solids. Early work by Argon and co-workers \cite{argon1979plastic} attempted to understand such rearrangements by considering disordered bubble rafts as tractable model systems for glasses. These investigations indicated that under shear deformation, plastic rearrangements involved the participation of a small number of bubbles. These localised rearrangements have been termed {\it shear transformations}, and have been subsequently observed in numerous computer simulations and experiments \cite{falk1998dynamics,Maloney2006,Schuh2007a,chikkadischall2012,jensen2014}. A schematic view, obtained from computer simulations of a two dimensional glass model \cite{falk1998dynamics} is shown in Figure \ref{fig:STZ}. Although one may envisage circumstances where plastic deformations may not be so localised, and will not occur as isolated events (see Ref. \cite{Nicolas2018} for a recent review and detailed discussion), the idea of localised shear transformations or shear transformation zones (STZ) underlies many lines of modeling plasticity. Unlike dislocations in crystals, such STZs do not have a well defined structural characterization, nor does one {\it a priori} have the ability to identify locations where plastic events will occur. Considerable effort has been directed therefore at arriving at ways of identifying the locations of plastic events, based on an analysis of local structure, energies, local yield stress, soft modes, {\it etc.} \cite{shi2007,ding2014,manningliu2011,patinet2016,Tyukodi2016c}, including the deployment of machine learning approaches \cite{cubuk2015,cubuk2016}. However, the possibility of predicting with high reliability the locations of plastic events remains a subject of ongoing exploration.

\begin{figure}[ht!]
\centering
  \includegraphics[scale=0.5]{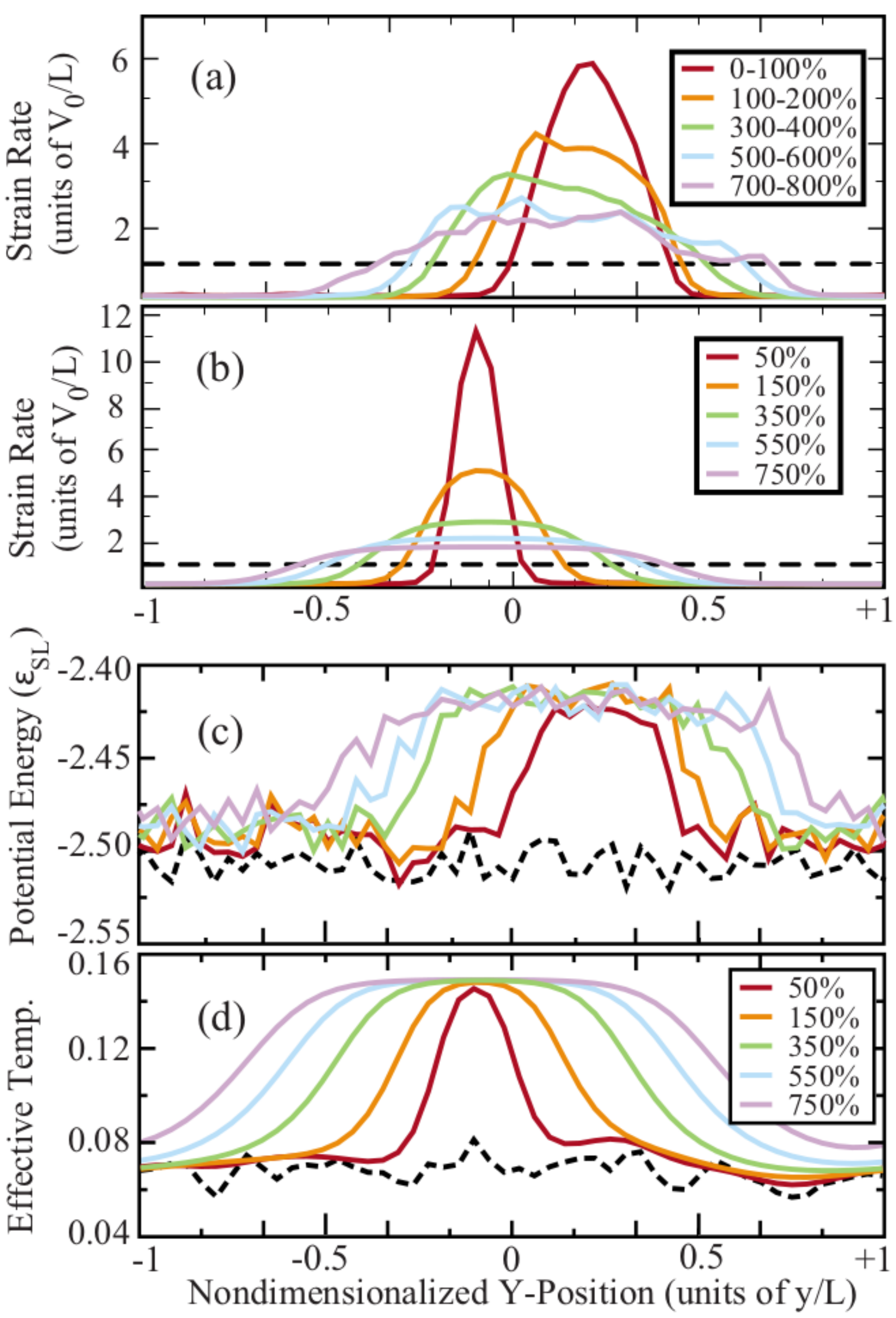}
  \caption{\color{black}{Comparison of simulation results and STZ theory. (a) Simulated position (Y) dependent strain rates, as functions of the scaled position for various strains, when the system is subjected to a global strain at a fixed rate. The dark dashed line is the imposed strain rate. (b) Corresponding estimates of the strain rates from the STZ theory. (c) Simulated potential energy per atom as a function of position for various strains. (d) The STZ prediction of effective temperature as a function of position. The dark dashed lines in panels (c) and (d) show the initial values for the potential energy and effective temperature. (source \cite{manningSTZbanding2007})}}
  \label{fig:STZbands}
\end{figure}

Based on the idea that STZs exist, may be created or destroyed, and can undergo transitions between different states in response to applied shear, Falk and Langer \cite{falk1998dynamics,Falk2011} proposed a description of plasticity termed STZ theory. In this description, the population of STZs is described to be governed by an effective or mechanical temperature $\chi$. Based on a nonequilibrium  thermodynamic formulation of the STZs \cite{bouchbinderetal,Falk2011}, expressions for the evolution of the STZ population, the rate of plastic deformation, and the effective temperature itself $\chi$ (which evolves during the deformation towards a steady state value governed by the shear rate) have been proposed. Yielding as a function of applied stress arises as a dynamical instability beyond a critical applied stress. 
In the original formulation, the STZ theory has no spatial dependence, but it does so in an extended form where the mechanical temperature is treated as varying as a function of space and time \cite{manningSTZbanding2007}. 
These analyses manage to capture many aspects of mechanical response in amorphous solids. An example is shown in Figure \ref{fig:STZbands}, which compares computer simulation results of the formation and spreading of shear bands with the predictions of STZ theory\cite{shi2007,manningSTZbanding2007}. In the mentioned work, the local average potential energies from simulations are compared with the effective temperatures. Although the notion of a mechanical temperature and the comparison with energies are both reasonable and well motivated, there is so far no concrete prescription for {\it measuring} this temperature. The other key notion, of an STZ population, with distinct states of occupation, also requires a concrete prescription to directly access it which is currently missing, and has indeed motivated some of the research to identify locations in a glass that are involved in plasticity \cite{shi2007,ding2014,manningliu2011,patinet2016}. 

% strain fields of STZs and Eshelby

\begin{figure}[ht!]
  \includegraphics[scale=0.33]{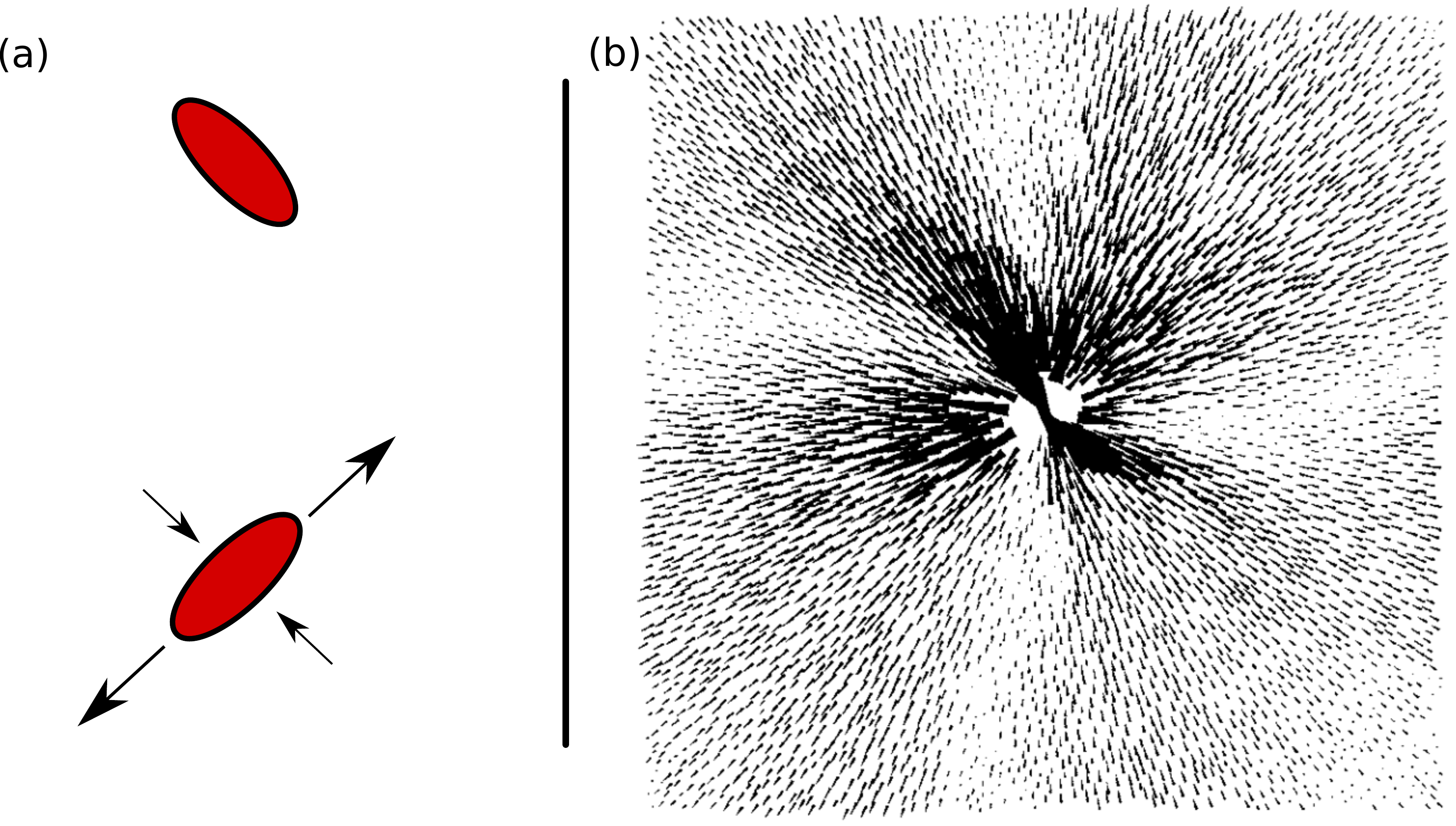}
  \caption{\color{black}{(a) Schematic representation of an Eshelby inclusion. (b) The tangential non-affine displacement field during a plastic event from computer simulations. (source \cite{Maloney2006})}}
  \label{fig:eshelby}
\end{figure}

%\begin{figure}[ht!]
%\centering
%  \includegraphics[scale=0.3]{DisplacementField.pdf}
%  \caption{\color{red}{The tangential non-affine displacement field during a plastic event from computer simulations. (source \cite{Maloney2006})}}
%  \label{fig:dispfield}
%\end{figure}

A more elaborate understanding of plastic events is provided by considering the nature of the perturbation they cause to the surrounding solid. Considering an idealisation of a point shear transformation, under other simplifying assumptions, Picard {\it et al} \cite{Picard2004} derived an expression for the displacement field (correspondingly the strain field, and assuming linear elasticity, the stress field) in two dimensions, which exhibits a {\it quadrupolar} form: 
\begin{equation} \nonumber
\sigma_{xy} ({\bf r}) \propto \int d{\bf r{'}} G({\bf r} - {\bf r{'}}) \epsilon^{pl}_{xy} ({\bf r{'}})
\end{equation} 
where $\epsilon^{pl}_{xy}$ is the plastic shear strain at some point, which causes an incremental stress $\sigma_{xy}$ elsewhere in the system, with the propagator $G$ given by
\begin{equation}\nonumber
    G(r, \theta) \propto { cos(4 \theta) \over r^2 }.
\end{equation}
This result carries over to three dimensions, with the propagator varying as $1/r^3$. The analysis by Picard {\it et al} is related to earlier work by Eshelby \cite{Eshelby1957}, on the elastic field caused by an {\it inclusion} in an elastic medium that undergoes a change of form, illustrated schematically in Figure \ref{fig:eshelby} (a). Signatures of non-affine displacements arising from a localised plastic event have been investigated in a number of numerical and experimental studies \cite{falk1998dynamics,Maloney2006,Schuh2007a,chikkadischall2012,jensen2014}. An early example is shown in Figure \ref{fig:eshelby} (b).  These analyses reveal that the effect of a localised plastic event is long ranged, bearing a characteristic anisotropic form, which has been found to be a valid description for amorphous solids. 

Such propagation of stresses constitutes a mechanism for long ranged and anisotropic interactions between past and prospective regions of plasticity, and has consequences for theoretical descriptions and modeling.  On the one hand, some investigators have argued that the long range nature of the interactions between plastic events implies that  mean field descriptions should have a high degree of validity for real systems \cite{Dahmen2009}. On the other hand, a class of models, termed elastoplastic models, have been developed taking cognizance of the explicit nature of stress propagation \cite{Nicolas2018}. We describe the general characteristics of elastoplastic models and indicate some developments based on their study first. 

% Elastoplastic models 

\begin{figure}[ht!]
\centering
  \includegraphics[scale=0.3]{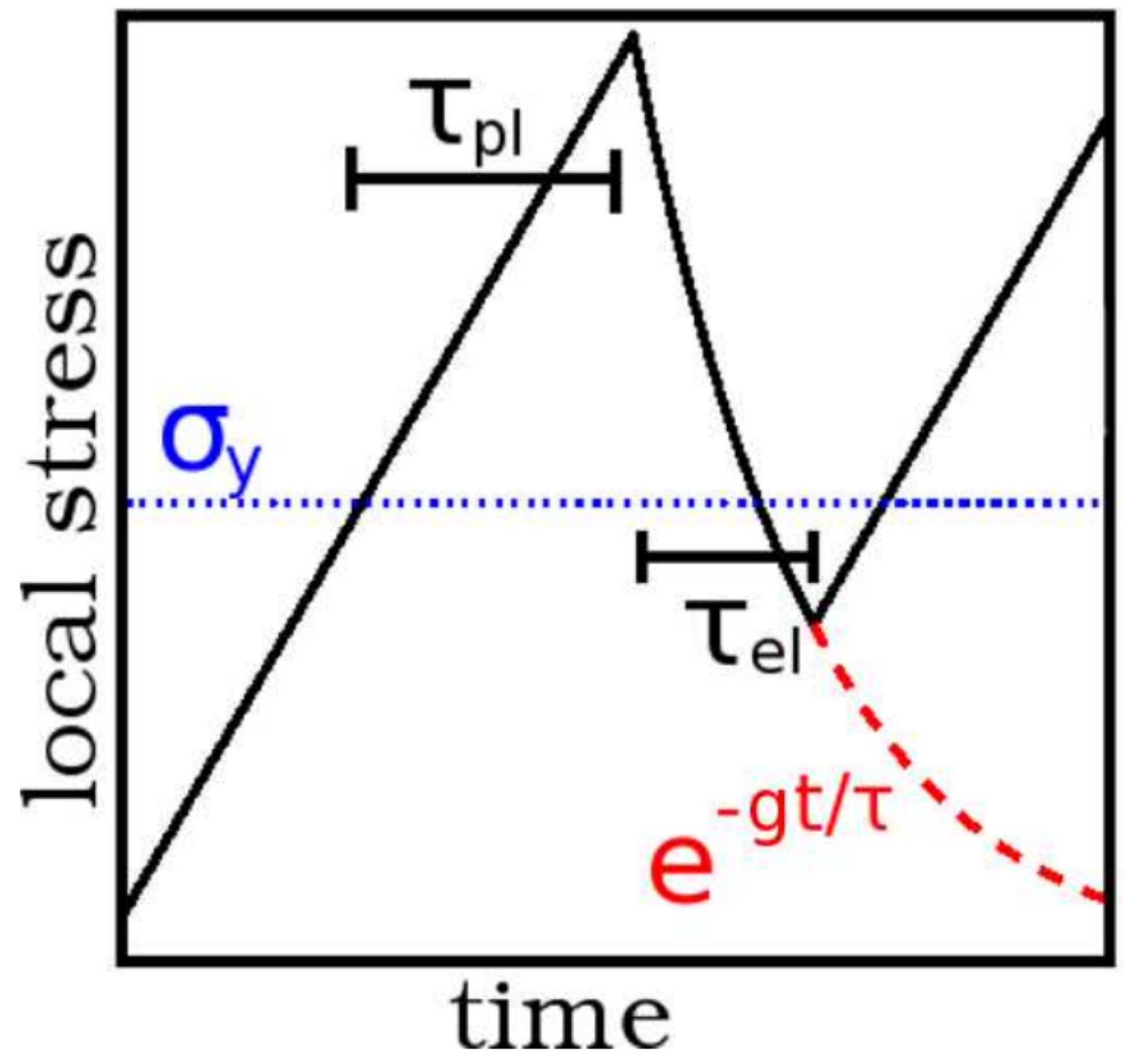}
 \caption{Schematic representation of the variation of local stress in an elastoplastic model. For  stresses larger than the threshold $\sigma_Y$, a site yields with a probability $1/\tau_{pl}$ (plastic rate). The site is active for a 
 a duration given by the restructuring time $\tau_{el}$. (source \cite{martens2012spontaneous})}
  \label{fig:EPScheme}
\end{figure}

Elastoplastic models represent an amorphous solid as being composed of cells, each of which may be thought of as representing a subvolume of the size of a plastic event. Each cell has a stress value, and a local yield stress value that is drawn from a distribution that forms part of the description of the model. A binary variable represents whether a plastic event is triggered at the cell or not. The stress at all sites increases monotonically with time with the increase of strain (typically at a fixed strain rate). A plastic event is triggered when the stress at a given site exceeds the local yield stress at that site. The stress then is propagated to other sites according to the propagator discussed earlier (or simpler ones) at a specified rate, and it relaxes at the site of the plastic events. Many variations of this basic strategy can be considered, including varying rates at which a plastic event is triggered, inclusion of thermal activation, {\it etc.} The expected variation of stress at a given site is shown in Figure \ref{fig:EPScheme}. In addition to the fully explicit models of the kind described, various works have also considered mean field versions, such as the Hebraud-Lequeux model.  This model  describes the time evolution of the distribution of local stresses, dependent on a parameter $\alpha$ that controls stress diffusion, in proportion to the number of sites that display plastic activity \cite{Hebraud1998}. One obtains a transition from Newtonian fluid behaviour at low shear rates to yield stress fluid behaviour obeying the Herschel-Bulkley constitutive law, as $\alpha$ is varied. Elastoplastic models have also been employed to study {\it avalanches}, correlated plastic events that occur in tandem, typically in the steady flow regime \cite{Talamali2012,Liu2016}. Avalanches in the steady state are relevant if one is interested in the dynamics of plasticity in that regime\cite{salerno2013} (for example, strain localisation, which we discuss in the next section), and has been of interest in the context of intermittent dynamics in earthquakes and other driven systems. 

% Depinning, RFIM 

The nature of avalanches approaching yielding have also been addressed by many studies, motivated by analogies with related systems. We mention two such cases, which will be referred to in discussion of recent results in the next section. Dahmen and co-workers \cite{Dahmen2009} have proposed a model for avalanches and yielding, motivated by ideas emerging from the study of the movement of interfaces in the presence of pinning sites, that predicts avalanches with a power law distribution with a cut off, with a universal power law exponent of $-3/2$, and a mean avalanche size that diverges as the yielding point is approached. While a variety of results have been compiled to show the validity of this description \cite{DahmenNphys2011,DahmenSciRep2015}, recent results described in the next section  raise questions on some aspects of such a description of yielding. Another model that has been used as a reference is the random field Ising model (RFIM), in which, in addition to the external field $H$, a random field is applied at each site. The strength of the random field $R$ is a key control parameter, and in the $H$-$R$ plane, the zero temperature phase diagram of this system exhibits a critical point at a critical disorder $R_c$, and at lower $R$, a spinodal limit defines the equivalent of the yielding point. Analysis in Ref. \cite{NandiBiroliTarjusPRL2016} indicates that at the transition point, the susceptibility (associated with avalanche size) does not diverge, unlike the mean field case. The implications of this picture to yielding in amorphous solids are discussed in the next section.

% Energy landscape picture 

Another perspective from which plasticity and yielding in glasses and amorphous solids has been analysed is to focus on the energy landscape properties of the deformed glasses. Such analysis has typically been conducted in the athermal quasi-static limit (AQS), in which the solids are studied in the zero temperature limit, subject to quasistatic strain, such that the glasses always reside in energy minima \cite{MalandroLacks1999,Lacks2004a,Karmakar2010,HentscheletalPRE2011,Dasgupta2012,dasgupta2013shear,Itamar2016yielding,parisi2017shear,lerner2018protocol}. Plastic rearrangements arise as a result of the loss of stability of the energy minimum that the system resides in, which constitutes a saddle-node bifurcation \cite{Karmakar2010,HentscheletalPRE2011}. Such analyses have been extensively conducted through computer simulations of atomic glasses and some results will be discussed in the next section. 

\section{Recent  Investigations} 
%10 pages 
%SS Cyclic deformation, annealing results, Itamar and co replica work, Wyart and co work on local yield (delta gamma), 

%\subsection{Cyclically Deformed Glasses}
%5-6 pages 

%\subsection{Other Results} 
% 4 pages 

% Distance to yield 

%ADSP: References relevant for avalanches: \cite{...}

% SS: Structure for this section: 
% Nature of the yielding transition, and avalanches (hard to separate unfortunately) -- diverging sizes expected from work of Dahmen, Wyart. Evidence for first order transition - cyclic shear, Procaccia work, well annealed glasses by Ozawa et al, Popovicc et al. Theoretical work by Zamponi et al. Simulation tests by Yuliang Jin. 
% Annealing effects (some already covered)
% Presence or absence of shear banding - cyclic shear, well annealed glasses, model studies with EP models, SGR. % Go ahead and organize whatever you have in this way. Don't worry about what you don't have, but if you can upload corresponding figures, that will be good. 

\begin{figure}[ht!]
\centering
\includegraphics[scale=0.3]{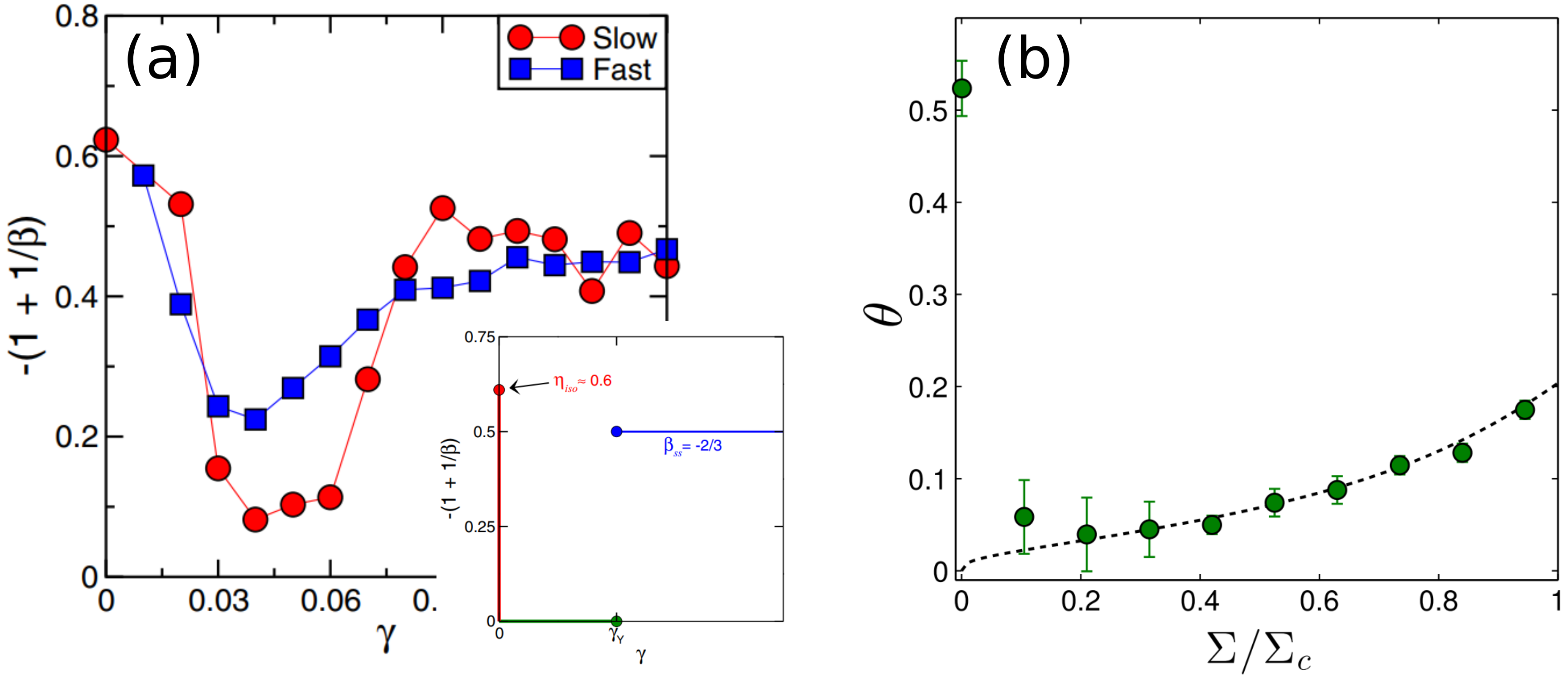}
 \caption{(a)  The exponent $\eta =-(1 + \frac{1}{\beta})$ for  different strain windows from an AQS simulation of a two dimensional atomic glass for fast and slowly quenched samples. The inset shows the theoretical expectation based on the analysis in \cite{HentscheletalPRE2015}
 (source \cite{HentscheletalPRE2015}). (b) The exponent $\theta$ ($\eta$) for an elastoplastic model exhibits similar behaviour and has been interpreted to 
 suggest criticality of avalanches at all strains below yielding. (source \cite{JinWyart2016}).}
  \label{fig:theta}
\end{figure}

A key question that has been addressed by many workers is the nature of the avalanches and how they may behave as the yielding transition is approached. We begin by a discussion of work in this regard, based on an energy landscape perspective. Procaccia and co-workers \cite{Karmakar2010,Lerner2009} have analysed the scaling of the stress drop $\Delta \sigma$ and energy drops $\Delta U$ during steady state avalanches, observing that they both scale with N, with $\Delta \sigma \sim N^\beta$, $\Delta U \sim N^\alpha$, with $\alpha - \beta = 1$, and it has been found that for both two and three dimensions, $\alpha = 1/3$, and $\beta = -2/3$. On the other hand, it was found that $\beta \approx -0.62$ for the first plastic event in a freshly prepared glass (which is not universal but appears to depend on the glass preparation), and is related to the behaviour at small values of the distribution of distances to the next plastic event, $\Delta \gamma$, which is generally expected to vary as $P(\Delta \gamma) \sim \Delta \gamma^\eta$.  Analysing the first plastic event of a freshly quenched glass, Karmakar {\it et al} \cite{Karmakar2010} argued (and showed) that the full distribution agrees well with the Weibull distribution, leading to a relationship $\beta = -1/(1 + \eta)$. Hentschel {\it et al} \cite{HentscheletalPRE2015} considered the statistics of intervals to yield in finite strain windows below the yielding point, and made the interesting observation that $\beta$ values, from $\beta = -0.62$ at $\gamma = 0$ to $\beta = -2/3$ for strain $\gamma$ above the yield strain, vary non-monotonically, approaching lower values in between (Figure \ref{fig:theta} (a)).  Based on the analysis of a Fokker-Planck equation for the evolution of the Hessian eigenvalues, they argued that for finite strain, $\eta = 0$, and if the relation  $\beta = -1/(1 + \eta)$ remains valid in this regime, one would expect $\beta = -1$, which would correspond to having avalanches of unit size all the way to the yielding transition. 

A similar analysis, on the basis of elasto-plastic models and scaling analysis was performed by Jin {\it et al} \cite{JinGuedreRossoWyartPRL2015,JinWyart2016}, who concluded that for a given value of $\eta$ ($\theta$ in their notation), the mean avalanche size would scale as $<S> \sim N^{\theta/(1+\theta)}$. In this case, exponent values are greater than zero and vary continuously with stress (Figure \ref{fig:theta} (b)), as in Ref. \cite{HentscheletalPRE2015}. The non-zero values of the observed exponent are argued to imply system-spanning avalanches for all stresses (strains) below yielding. Similar observations are made by Ozawa {\it et al} \cite{Ozawa2018a} (discussed further below) albeit with the variation that with annealing of the glasses, the exponent $\theta$ approaches zero in the pre-yield regime.  In attempting to find a reconciliation, Lerner {\it et al} \cite{lerner2018protocol} have suggested that system sizes may matter strongly in accessing the relevant regime of the distributions of distances to local yield, infeasible at the present.

% Nature of the yielding transition

\begin{figure}[ht!]
\centering
  \includegraphics[scale=0.18]{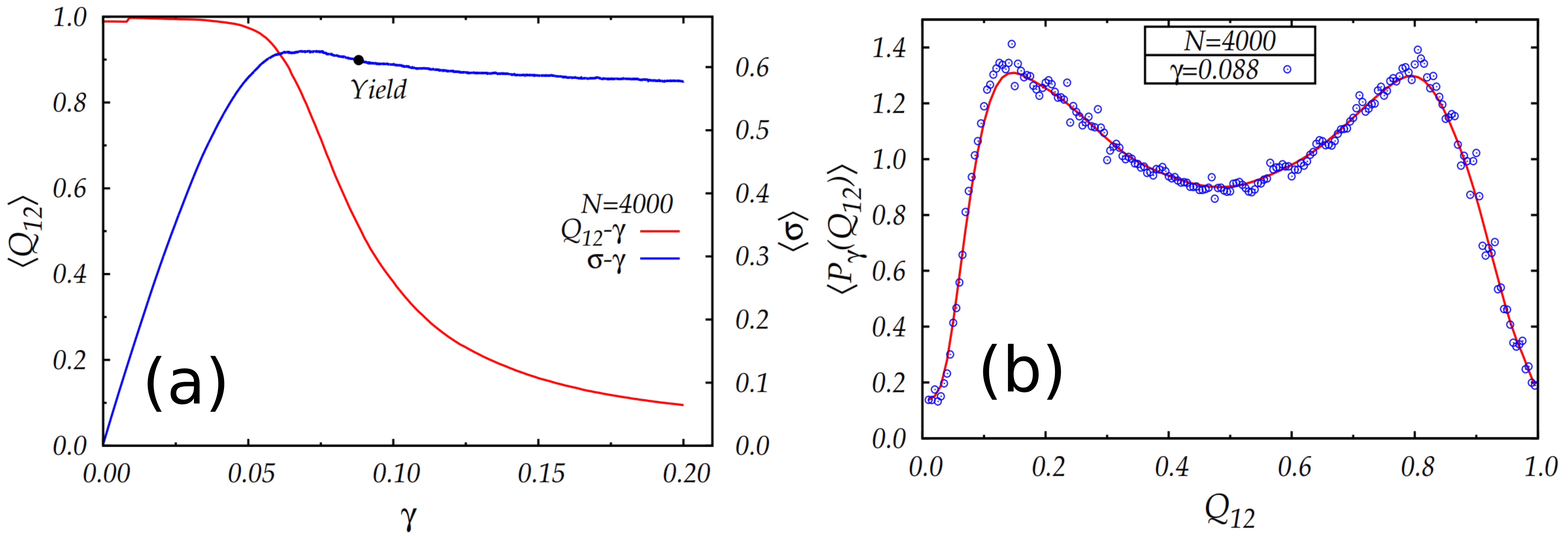}
 \caption{\color{black}{(a) Averaged stress {\it vs.} strain shown along with the overlap function $<Q_{12}>$ for the same of strain $\gamma$ (source \cite{Itamar2016yielding}). (b) The averaged probability distribution function $P_\gamma(Q_{12})$ at $\gamma_y = 0.088$ . At the ``yield" strain the distribution has two peaks with equal weights.}}
  \label{fig:overlap}
\end{figure}

We next discuss briefly some developments in discussing the nature of the yielding transition, before returning to the question of avalanches. 
A number of recent studies have addressed the nature of the yielding transition, with an emerging consensus that one must understand it as a spinodal limit. Based on  a mean field infinite dimensional description of the glass transition for hard spheres, yielding has been described as a spinodal point \cite{RainoneetalPRL2015,Urbani2017b}, and subsequent simulation studies have worked out some of the details of these calculations for finite dimensional systems \cite{Jin}.  Jaiswal {\it et al} \cite{Itamar2016yielding} explored the idea that yielding was associated with a delocalisation of glass configurations in configuration space. They considered an ensemble of initially nearby configurations, and computed the distribution of overlaps between them as a function of the imposed strain. This distribution is initially peaked at large overlaps, but becomes bimodal as the strain is increased. Yielding is identified as a {\it first order} transition, associated with the low overlap peak of the distribution becoming larger in weight (Figure \ref{fig:overlap}). This formulation is familiar in replica theories of the glass transition, and in Ref. \cite{parisi2017shear,procacciaetal2017}, a replica analysis of this problem is carried out, with a description of yielding as a spinodal instability, and is associated with the formation of shear bands. Application of random first order transition theory (RFOT) to the strength of glasses and shear banding has also been discussed by Wolynes and co-workers \cite{wisitsorasak2012strength,wolynesshearbanding}. To understand the origins of shear bands, Dasgupta {\it et al}\cite{Dasgupta2012,dasgupta2013shear} considered the energetics of arrangements of Eshelby-like inclusions in two dimensions, and showed that an organization of a large density of aligned inclusions along a line/plane become lower energy solutions than an inclusion-free solid, discontinuously at a yielding density. Such a calculation is informative, but does not provide a full analysis of under what conditions a shear banding instability may be observed in an amorphous solid. Shear bands form in some sheared glasses but not all. The evidence for the presence of avalanches that grow in size, or are system spanning, as yielding is approached is correspondingly varied. Some investigations that address and clarify these points are discussed below. 

%SS: Add more references to cyclic shear whereever suitable. Look at earlier papers. 

\begin{figure}
  \centering
  \includegraphics[scale=0.255]{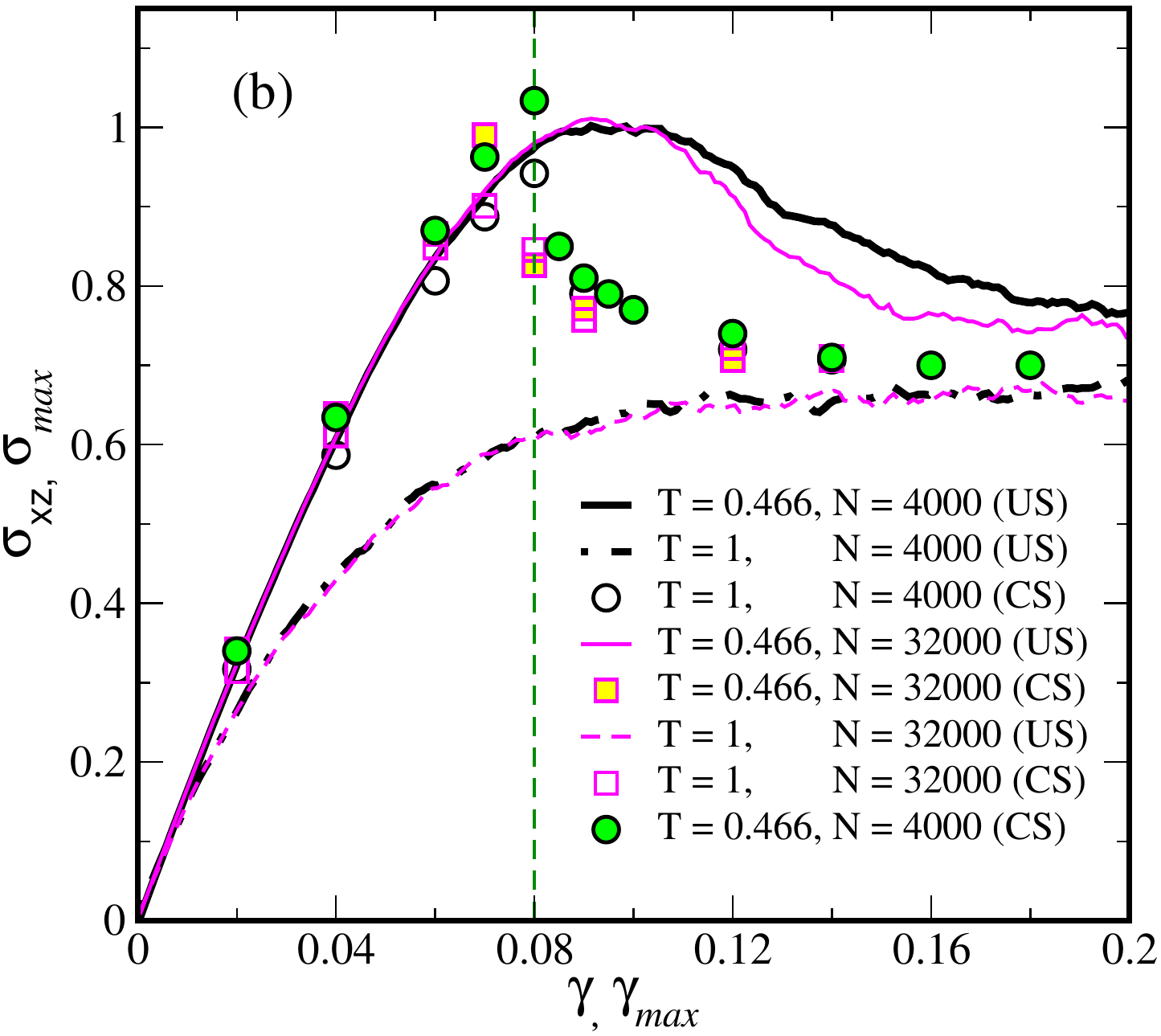}
  \includegraphics[scale=0.255]{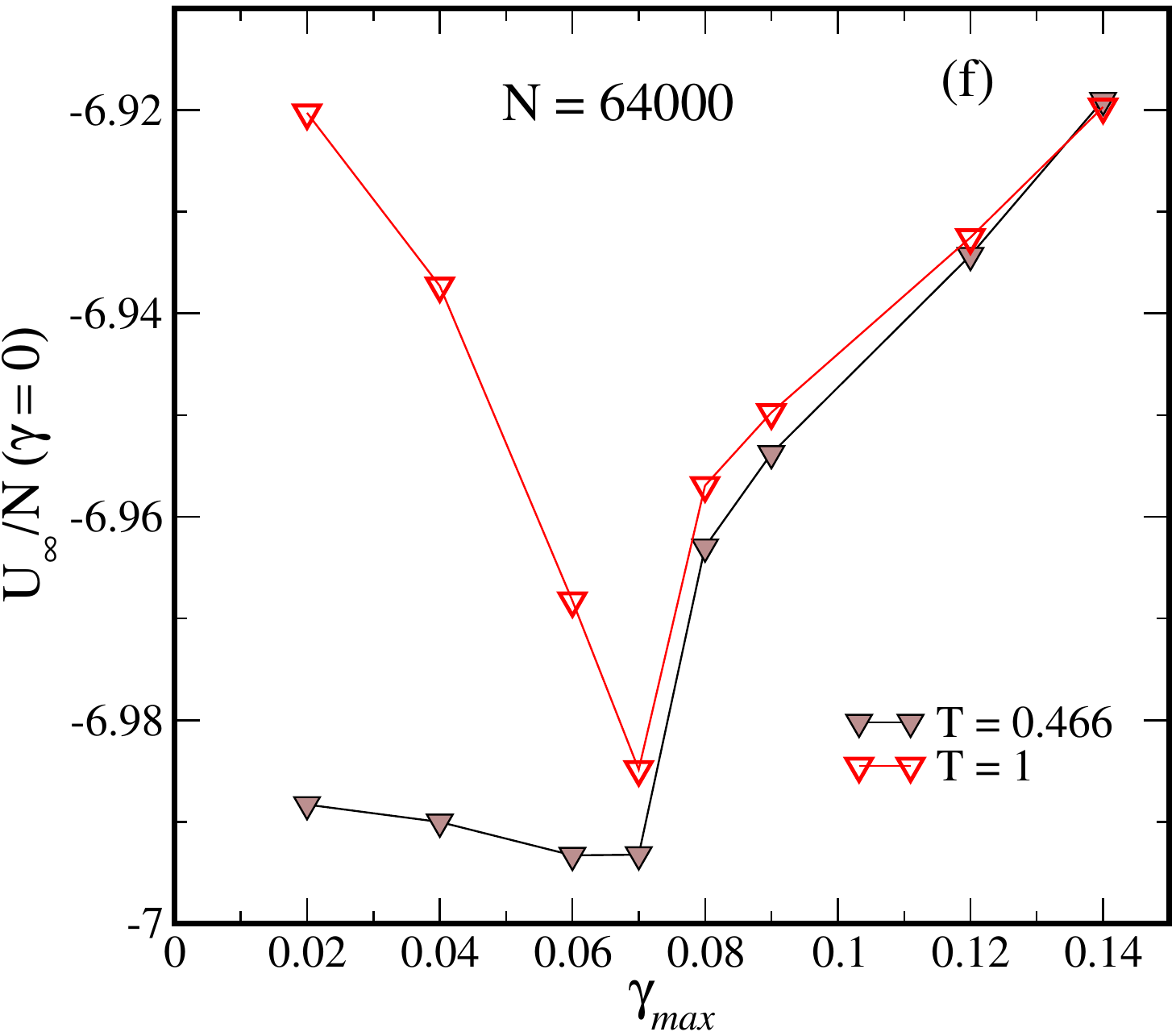}
  \includegraphics[scale=0.255]{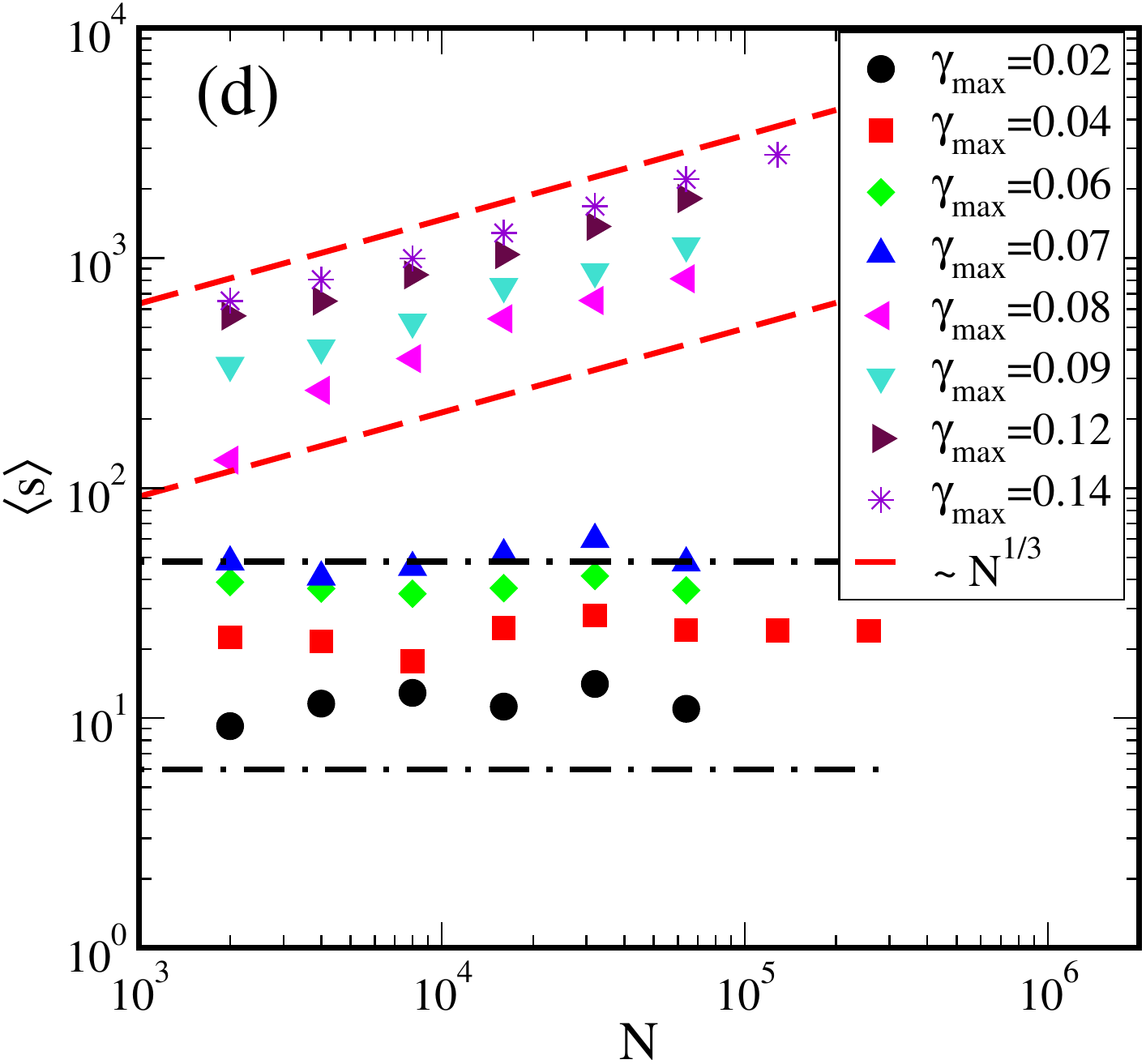}
  \caption{(a) Averaged stress-strain curves for uniform shear and cyclic shear
  for a 3D model glass (Kob-Andersen BMLJ), instantly quenched from the liquid at two different temperatures,  $T=1.00$ and $T=0.466$. Maximum stress $\sigma_{max}$ versus $\gamma_{max}$ is shown for cyclic shear. The vertical line at $g_{max}=0.08$ indicates the sharp yielding transition observed.
   (b) Asymptotic energy per particle at $\gamma=0$ plotted against the corresponding strain amplitude. Energies decrease with $\gamma_{max}$ until the
yield strain amplitude is reached ($\gamma_{max}\le \gamma_{y}$), after which they increase with $\gamma_{max}$. (c) Mean avalanche size versus system size $N$ shows no significant system size dependence for $\gamma_{max}\le \gamma_y$ but a clear $N^{1/3}$ dependence above. The discontinuous change in behaviour marks the yielding  transition, seen at $\gamma_{max}=0.08$. (source \cite{leishangthem2017yielding})}
  \label{fig:cyclic}
\end{figure}

Many investigations of mechanical response have been conducted employing cyclic or oscillatory deformation of glasses, either under the AQS protocol, or at finite rates and/or temperatures \cite{Lacks2004a,Fiocco2013,regev2013,PRIEZJEV2013,regev2015reversibility,leishangthem2017yielding,das2018annealing,PRIEZJEV2018,parmar2019strain}. Under cyclic deformation, with the strain amplitude as the control parameter, it is found that the glasses evolve in properties, but in distinct ways above and below a threshold amplitude that is identified as the yielding amplitude \cite{Fiocco2013,leishangthem2017yielding,parmar2019strain}. Glasses anneal ({\it i.e.}, reach lower energy states) with repeated cycles below the yielding strain, with the degree of annealing being the highest at the yielding transition (Figure \ref{fig:cyclic}). Glasses with different degrees of annealing behave very differently with uniform shear (poorly annealed glasses show a monotonic increase in stress {\it vs.} strain, whereas better annealed glasses show a stress overshoot), but considering the maximum stress as a function of the strain amplitude $\gamma_{max}$, a very different picture emerges. The stress strain curves show a discontinuous jump at the yielding amplitude. The nature of avalanches below and above yielding are qualitatively different. Avalanches below yielding are finite, and do not grow upon approach to the yielding point, or with system size, whereas avalanches above yielding are system spanning, and show the $N^{1/3}$ scaling previously described. Thus, the conclusion from this work is that yielding occurs as a discontinuous transition, and avalanche sizes below yielding show no indication of growth upon approaching the yielding point, nor with system size. The nature of the yielding transition in cyclic shear is strongly influenced by the annealing of the glasses that is observed.  Cyclic deformation has been employed, in addition to studies of yielding, in understanding transitions from reversible to irreversible behaviour\cite{pine2005chaos,corte2008random}, and memory formation \cite{keimMemoryReview2018}, which we do not discuss here, but details of which may be found in Ref. \cite{keimMemoryReview2018}.

\begin{figure}
  \centering
  \includegraphics[scale=0.15]{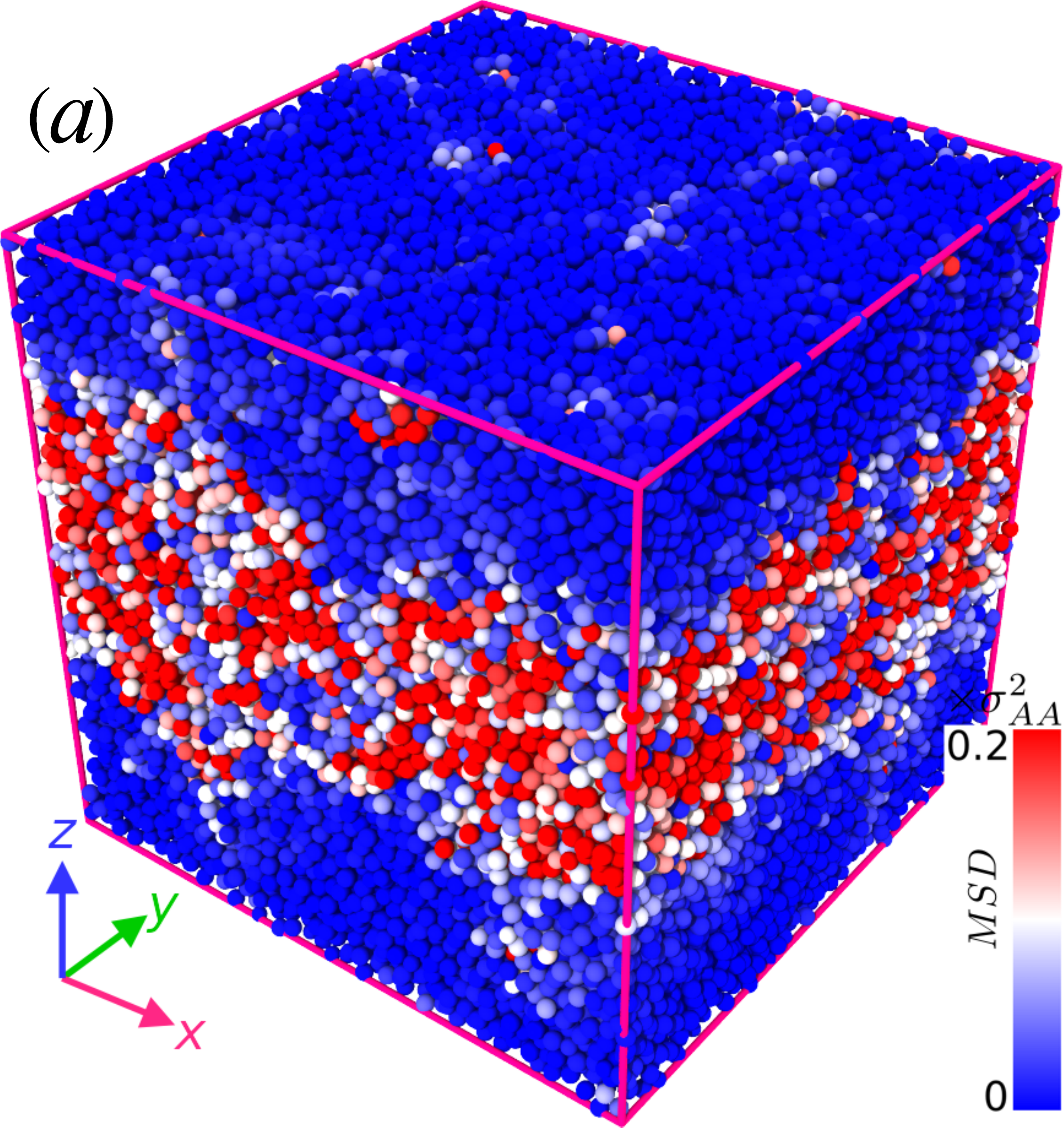}
  \includegraphics[scale=0.25]{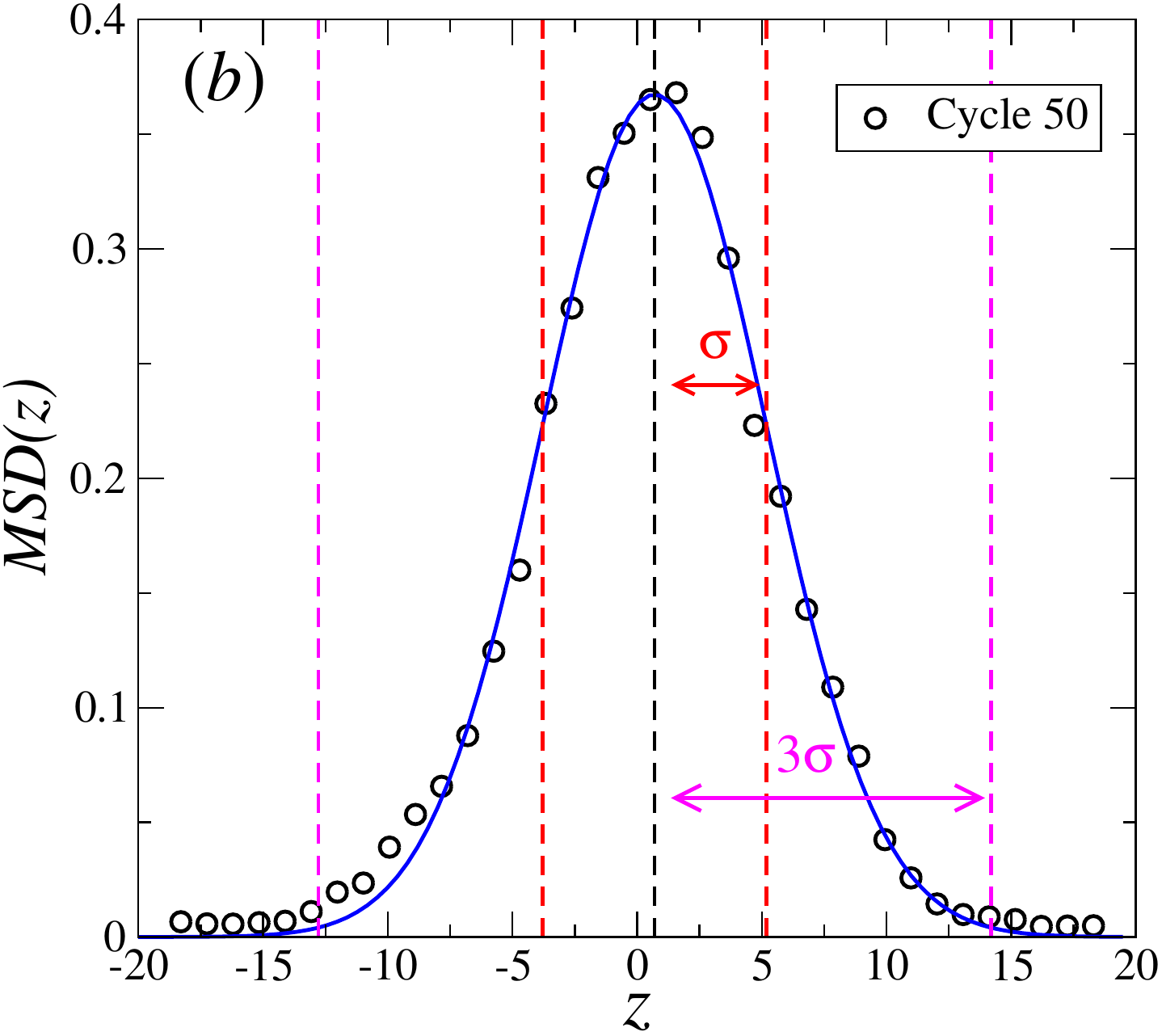}
  \includegraphics[scale=0.25]{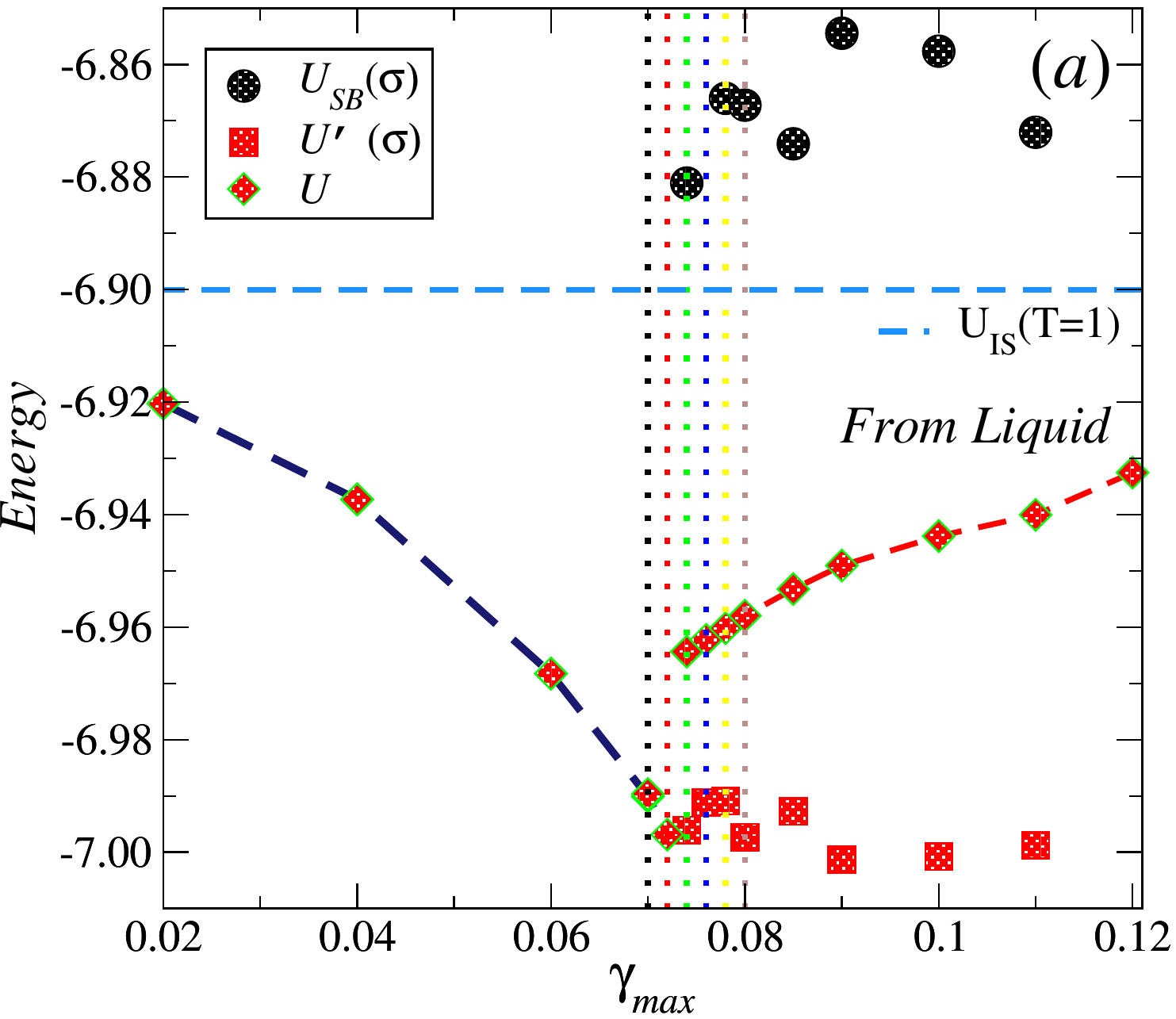}
  \caption{(a) A sample configuration of particles from the steady state, for a cyclially sheared glass, at  strain amplitude $\gamma_{max}=0.09$ above the yield strain.  The color map is based on the displacement of particles between two consecutive cycles at zero strain. Highly mobile particles (Particles with squared displacement $ > 0.2 \sigma^{2}_{AA}$) are colored in red, whereas particles in blue move considerably less. (b) Mean squared displacement between two cycles is shown as a function of the coordinate $z$ in the shear direction, along with a Gaussian fit.  (c) The mean energy changes discontinuously across the yielding transition. The subsystem outside the shear band is annealed for amplitudes even beyond the yield amplitude, whereas particles within the shear band have higher energies (source \cite{parmar2019strain}).}
  \label{fig:shearband}
\end{figure}

For amplitudes larger than the yielding amplitude, the cyclically sheared glasses exhibit strain localisation, that is stable in characteristics for a given amplitude (the width, for example) but some interesting features are observed (Figure \ref{fig:shearband}) -- The energies of the particles within the shear band become the highest possible ({\it top of the landscape}, as discussed in the context of inherent structures (energy minima) in glass forming liquids \cite{Sastry1998c,angellNandV1998,utz2000}) while the energies outside the shear band continue to decrease with increase of strain amplitude, and display logarithmic relaxation in cycle number as also seen in liquid simulations \cite{parmar2019strain,das2018annealing} and for homogeneous glasses below the yielding amplitude. These observations are relevant for our discussion of shear banding and strain localisation below. 

\begin{figure}
  \centering
  \includegraphics[scale=0.35]{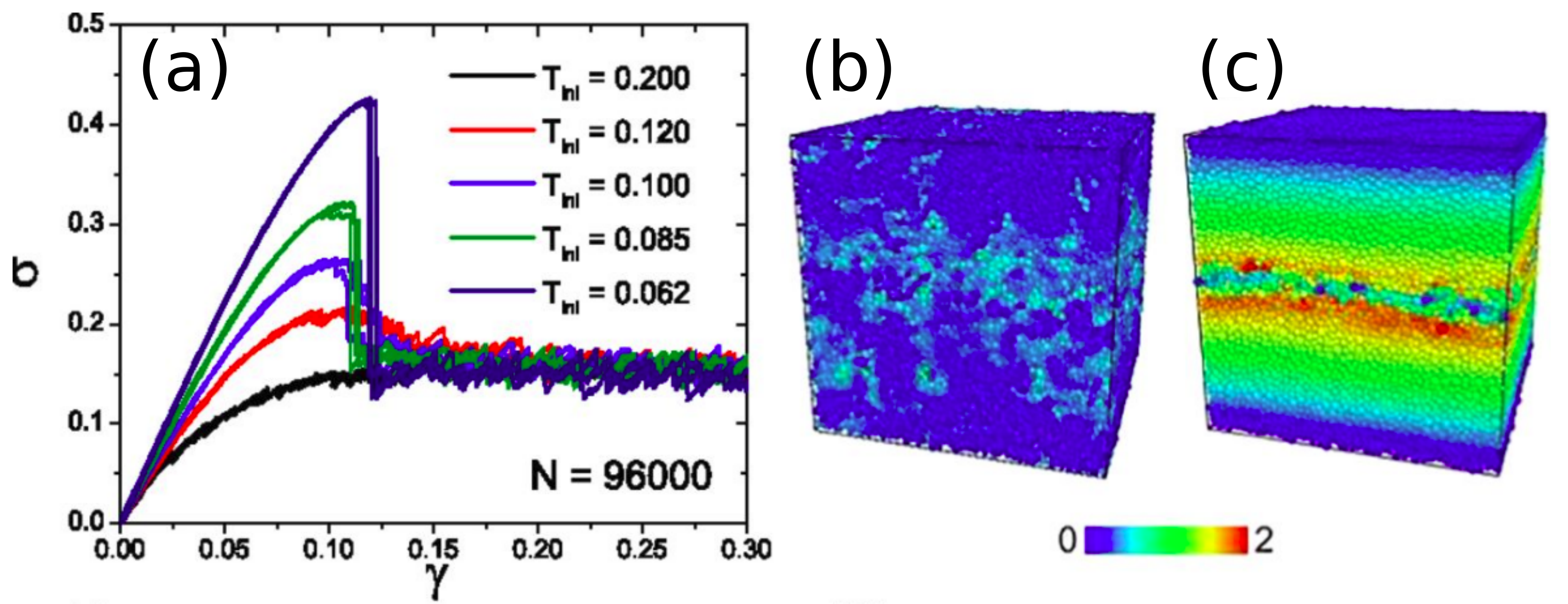}
  \caption{(a) The stress-strain curves for samples prepared at several temperatures ($T_{ini}$), which determine their degree of annealing. The role of annealing in the response is evident, with larger discontinuous jumps with higher annealing, below a critical $T_{ini}$, above which the stress-strain curves are continuous. 
   (b \& c) Snapshots of the system at the yield point for cases with continuous (b) and discontinuous (c) stress-strain curves. The color code is based on  non-affine displacements between the initial configuration and the configuration at the yield strain. (source \cite{Ozawa2018a}).}
  \label{fig:Ozawa}
\end{figure}

The implications of annealing on the nature of yielding has been considered by Ozawa {\it et al} \cite{Ozawa2018a}, who considered glasses that were annealed to different degrees, obtained by equilibrating a model liquid over a wide range of temperatures, and quenching them rapidly to zero temperature. They find that for low enough initial temperatures (or degree of annealing), the glasses exhibit yielding associated with a sudden drop in stress. Such sudden stress jumps are associated with strong strain localisation. But as the degree of annealing reduces, a more gradual transition with and without stress overshoots is found (Figure \ref{fig:Ozawa}). These authors interpret their results in terms of the behaviour seen in the random field Ising model (providing evidence for such a comparison), and argue that the spectrum of yielding behaviours observed can be rationalised in terms of the zero temperature phase diagram of the RFIM, with  a random critical point separating brittle and ductile yielding behaviours. 

Ozawa {\it et al} \cite{Ozawa2018a} also analyse a mean field elastoplastic model, paying particular attention to the initial distribution of stresses that captures the degree of annealing. In such a calculation also, the trend observed with annealing above is recaptured. A very similar investigation is also found in Ref. \cite{Popovic2018a} who draw similar conclusions. It is interesting that the mean field model in \cite{Dahmen2009} with {\it weakening} produces similar results. A somewhat intriguing feature of all these analyses is that they capture, at the mean field level, behaviour that is supposed to be associated with strong spatial correlations (strain localisation), and is also expected to arise from a specific form of the stress redistribution.  

\begin{figure}
  \centering
  \includegraphics[scale=0.23]{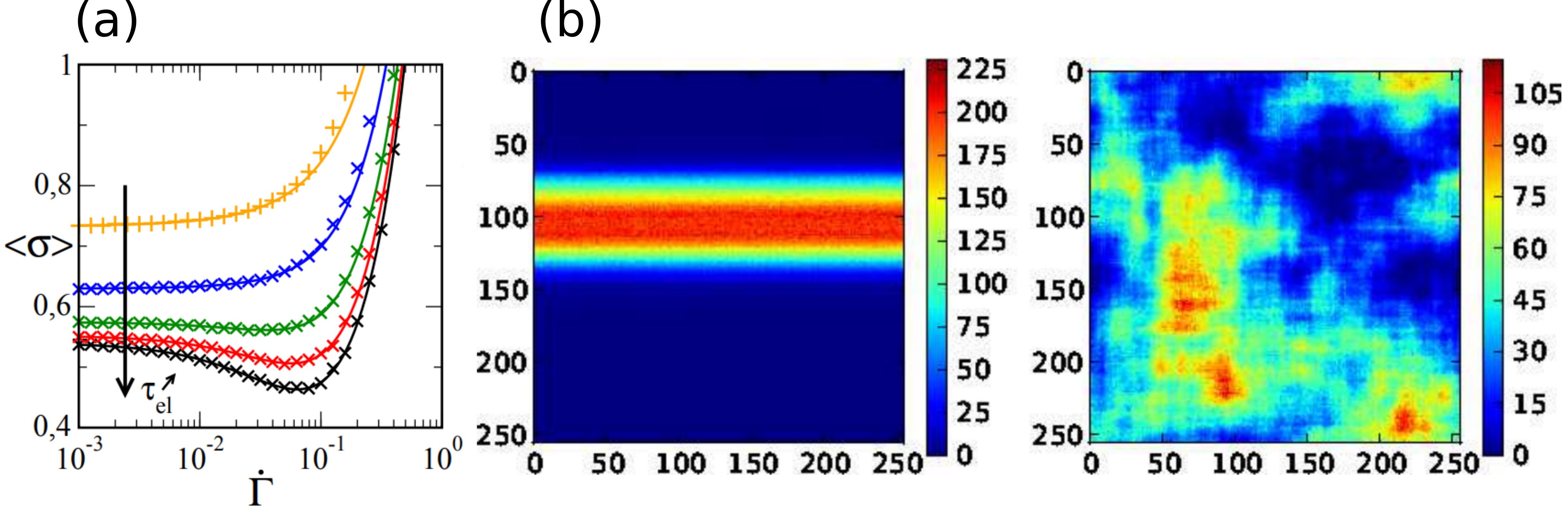}
  \caption{(a) Flow curves from a mean field calculation, and (b) shear band formation or otherwise in an elasto-plastic model simulation employing an Eshelby-type vs short range propagator. (source \cite{martens2012spontaneous})}
  \label{fig:Martens}
\end{figure}

The presence of shear bands in flow or strain localisation have been studied extensively in the past, both experimentally and in computer simulations, in soft glassy materials and hard glasses, as described in Ref.  \cite{divoux2010transient,moorcroft2011age,moorcroft2013criteria,fielding2014shear,divoux2016shear,Bonn2017c,chaudhuri2013onset,shrivastav2016yielding,vasisht2017emergence,Nicolas2018}. In the context of hard glasses, strain localisation has been observed and discussed in the context of yielding, as already mentioned several times. Apart from the work mentioned above, there have been some observations, starting from \cite{shi2007}, that a correlation exists with the annealing of glasses and strain localisation during yield. However, a physical understanding of the connection is only now beginning to emerge. On the other hand, one may also inquire as to whether the strain localisation observed at yielding will persist or the shear band will grow and engulf the system as the strain increases. In the rheological context of mainly soft glasses, the persistence or otherwise of shear bands, and dependence on shear rates and other parameters, have been analysed by a variety of investigations  \cite{jagla2007strain,coussot2010physical,jagla2010shear,martens2012spontaneous,Radhakrishnan2016b,shrivastav2016yielding,Falk2018shear,vasisht2017emergence,vasisht2018permanent}. The physical picture of the theoretical analyses invoke some form of  competition between rates of {\it aging}, which is treated as a spontaneous process, and {\it rejuvenation}, which is often associated with shearing. Following Coussot and Ovarlez \cite{coussot2010physical}, Martens {\it et al} \cite{martens2012spontaneous} analysed a related mechanism, where the relaxation time taken by an activated region to relax stresses plays a key role (refer to Figure \ref{fig:EPScheme}). When this time scale $\tau_{el}$ gets long enough, permanent shear bands arise. In a mean field calculation, this can be seen in non-monotonic flow curves (Figure \ref{fig:Martens}) and in a simulation, in the form of localisation of shear. In the latter case, shear banding is observed only when an Eshelby-type stress propagator is used but not when a short range propagator is used. This latter observation seems to support the idea that the nature of stress propagation is of key importance, as in the analysis in Ref.~\cite{Dasgupta2012,dasgupta2013shear}. But once again, mean field or other abstract models \cite{martens2012spontaneous,Radhakrishnan2016b} appear to capture shear banding even without explicit consideration of the nature of stress propagation in space. Further, aging and rejuvenation rates are not available when one considers AQS dynamics, and where the effects of annealing appear in the rheological models is also not clear. In the context of cyclic shear discussed above, even though performed under AQS conditions, aging and rejuvenation effects are present and play a role, but a key difference from some of the model studies mentioned is that both aging and rejuvenation are induced by the shear deformation itself, heterogeneously in space. It would be interesting to have theoretical analyses that will encompass these scenarios. 

\section{Outlook} 

The discussion in the previous two sections clearly indicates that much progress has been made in recent times in answering some of the key questions related to yielding in amorphous solids, concerning the nature of elementary processes, their interactions,  the nature of the yielding transition, the dependence of the mechanical response on the preparation history and structural properties of the amorphous solids, and the presence or otherwise of strain localisation and shear banding during yielding and flow. There is a convergence of different approaches that at the outset are not very related ({\it e. g.}, elasto-plastic models vs. glass transition theory), and increasingly, a convergence of what the nature of the yielding transition is. The degree of annealing of glasses has been seen as a key factor, and for well annealed glasses, a sudden, discontinuous yielding transition without precursors appears to be the commonly observed behaviour. Annealing is expected to change the structure, but whether one can structurally identify locations where plastic deformation events can take place is an interesting subject of ongoing investigation. A discontinuous yielding transition resulting from a spinodal instability have been arrived at in a variety of ways. But the nature of the spinodal (whether critical, not, and what kind of divergences may be expected, if any) will continue to be investigated in the near future, but the questions are sharper than they were before. There isn't yet fully clarity or consensus on the question of the nature of avalanches before yielding, in particular whether they would be system spanning or not. The origins, nature and role of aging and rejuvenation effects is another area where open questions remain to be answered. As described, cyclic shear offers an interesting avenue to investigate key aspects of yielding but current modeling approaches are limited in this regard. For example, elastoplastic models are rule based {\it automata}, which do not easily permit the notion of reversing the strain. Energy or free energy based formulations, such as those of Jagla \cite{jagla2007strain,jagla2010shear,jagla2017} are very interesting in this context to explore as starting points. At finite shear rates and finite temperatures, the location and nature of yielding can change substantially ({\it e. g.} \cite{parmar2019strain}). Systematically revisiting those features will be important after the athermal limit is well understood. 

After this survey of theoretical approaches, it may well be appropriate to end by returning to practical matters, and ask if these ideas will in anyway ``help make better glass", paraphrasing P. W. Anderson \cite{AndersonPerspective}. Indeed, a key feature of the behaviour of well annealed glasses is brittle failure, which is not a desirable feature, while other properties of such annealed glasses are. Are there ideas for how such failure can be avoided? Based on the understanding of the nature of the yielding instability, various investigators have proposed the inclusion of pinning centers and micro-alloying as a way of curtailing run away instabilities \cite{Microalloying,metallicGlassLiterature1,metallicGlassLiterature2,metallicGlassLiterature3,metallicGlassLiterature4,bhowmik2018effect,bhowmik2019particle} and thereby make annealed glasses more resistant to brittle failure, and more ductile. More such insights may arise from the development of a unified view of yielding and flow of amorphous solids in the coming years. 

\bibliographystyle{apsrev4-1}
\bibliography{parsas}

\end{document}